\documentclass[reprint,amsmath,amssymb,iop]{revtex4-1}
\usepackage[T1]{fontenc}
\usepackage{color}
\usepackage{graphicx}
\usepackage{epstopdf}
\usepackage{bm}
\usepackage{dcolumn}

\def\mc {\mathcal}

\newtheorem{fact}{Fact}
\newcommand{\tr}{\mbox{Tr}}
\newcommand{\bra}[1]{\mbox{$\langle #1 |$}}
\newcommand{\ket}[1]{\mbox{$| #1 \rangle$}}

\newcommand{\kb}[2]{\ensuremath{| #1 \rangle\!\langle #2 |}}

\pacs{
FILL
}

\begin{document}
\title{Asymptotic properties of entanglement polytopes for large number of qubits}

\author{Tomasz Maci\k{a}\.{z}ek$^1$ and Adam Sawicki$^{1}$}
\email[E-mail: ]{maciazek@cft.edu.pl, a.sawicki@cft.edu.pl}
\affiliation{$^1$Center for Theoretical Physics PAS, Al. Lotnik\'ow 32/46, 02-668, Warsaw, Poland}

\begin{abstract}
Entanglement polytopes have been recently proposed as a way of witnessing the SLOCC multipartite  entanglement classes using single particle information. We present first asymptotic results concerning feasibility of this approach for large number of qubits. In particular, we show that entanglement polytopes of $L$-qubit system accumulate in the distance $O(\frac{1}{\sqrt{L}})$ from the point corresponding to the maximally mixed reduced one-qubit density matrices. This implies existence of a possibly large region where many entanglement polytopes overlap, i.e where the witnessing power of entanglement polytopes is weak. Moreover, we argue that the witnessing power  cannot be strengthened by any entanglement distillation protocol, as for large $L$ the required purity is above current capability. 
\end{abstract}

\maketitle
\section{Introduction}
Correlations in multipartite quantum systems possess surprising and useful properties that have been a subject of studies since the very beginning of quantum theory \cite{Horodecki}. A pure state $\psi\in \mathcal{H}$ of a quantum  system consisting of $L$ subsystems is not entangled (separable) if and only if it is a simple tensor, i.e. $\psi=\psi_1\otimes\psi_2\otimes\dots\otimes \psi_L$. The standard approach to the classification of entanglement is based on the fact that quantum correlations remain unchanged under the action of local transformations called Local Operations and Classical Communication (LOCC) \cite{Nilsen}. Two states have the same type of entanglement if they can be transformed to each other using LOCC transformations. Recently, the set of maximally entangled states under LOCC operations in $L$ qubit systems has been determined \cite{KrausLOCC}. Relaxing the LOCC constraint to - transformed to each other using LOCC transformations with a finite probability - we obtain the class of operations called  SLOCC (Stochastic LOCC). The classification of entanglement using SLOCC operations has been extensively studied for small numbers of distinguishable and indistinguishable particles (see for example \cite{Dur, Verstraete, Levay,kus,holweck, wallach1, Dokovic}). Extension of these results to many particles is, however, considered to be difficult as one needs to know the ring of all SLOCC invariants which in general is difficult to find \cite{wallach2, turner}. 

In this paper we focus on many qubit systems exploring recently proposed concept of entanglement polytopes (EP) \cite{walter,sawicki}. Mathematically, $L$-qubit SLOCC operations can be represented by the group $G=SL(2,\mathbb{C})^{\times L}$ (collections of $2\times 2$ invertible complex matrices with determinant one). The action of $g=g_1\otimes \ldots \otimes g_L\in G$ on the Hilbert space of $L$-qubit pure states $\mathcal{H}$ is defined by 
\[
g.\phi=\frac{(g_1\otimes\ldots\otimes g_L)\phi}{\|(g_1\otimes\ldots\otimes g_L)\phi\|},\,\,g_k\in SL(2,\mathbb{C}),\,\, \phi\in \mathcal{H}.
\]
The space of pure states is divided by this action into SLOCC entanglement classes:
\begin{eqnarray}
\mathcal{C}_\phi:=G.\phi=\{g.\phi:g\in G\}.
\end{eqnarray}  
Topological closures, $\overline{\mathcal{C}}$,  of classes $\mathcal{C}$ are not disjoint as they all contain the class of separable states. Moreover, typically for multipartite systems, there is an infinite number of SLOCC classes and the number of parameters distinguishing these classes grows exponentially with the number of particles. For qubits, only one-, two- and three- qubit systems have a finite number of SLOCC classes \cite{Dur}. Thus, the task of checking to which class a given state belongs, requires the knowledge that is typically not accessible experimentally. Taking into account these limitations, recently, a new approach for witnessing the SLOCC multipartite  entanglement classes has been proposed \cite{walter,sawicki}. It uses single particle information contained in the spectra of the one-particle reduced density matrices (RDM) in order to witness multipartite entanglement. The idea is based on certain algebraic and symplectic geometry results concerning the action of  groups on algebraic varieties and the associated with it momentum map that captures information about symmetries \cite{kirwan, ness}. The momentum map approach has proved to be a useful tool in quantum information theory and related fields. One of the main applications is the study of quantum entanglement, see \cite{sawicki,sawicki2,symplecticreview,SK11,SWK13,CCCQ,ST13,HKS13,maciazek1}. Momentum map techniques have also been used to solve the one-body quantum marginal problem \cite{klyaczko2,walter2}, whose solution in the context of fermionic systems gives rise to the generalised Pauli constraints and quantum pinning \cite{tennie2016pinning1,tennie2016pinning2,tennie2017influence,schilling2013natural,schilling2015hubbard,benavidespinning2,S15,OS17}. In the setting of this paper the momentum map results from the action of the local unitary operations $SU(2)^{\times L}$ on the space of pure states. It assigns to every $\phi\in \mathcal{H}$ the collection of the shifted RDMs, $\left(\rho_1(\phi)-\frac{1}{2}I,\ldots,\rho_L(\phi)-\frac{1}{2}I\right)$, where $\rho_i(\phi)$ is the $i$-th one-qubit RDM, i.e. 
\[\rho_i(\phi)=\frac{1}{\|\phi\|^2}\mathrm{tr}_ {1,\ldots,\hat{i},\ldots,L}\kb{\phi}{\phi}.\]
Next, for a state $\phi$  let $\left(p_i(\phi),1-p_i(\phi)\right)$ be the increasingly ordered spectrum of the one-particle RDM $\rho_i(\phi)$, where $p_i(\phi)\in[0,\frac{1}{2}]$.  The shifted local spectra, i.e. the spectra of
$\rho_{i}(\phi)-\frac{1}{2}I$ are given by $\{-\lambda_{i}(\phi),\lambda_{i}(\phi)\}$, where $0\leq\lambda_{i}(\phi)=\frac{1}{2}-p_{i}(\phi)\leq\frac{1}{2}$. The map that plays an important role is the map $\Psi$ defined by
\begin{eqnarray}\label{Psi}
\Psi:\mathcal{H}\rightarrow \left[0,\frac{1}{2}\right]^{\times L},\,\,\,\Psi(\phi)=\left(\lambda_1(\phi),\ldots,\lambda_L(\phi)\right).
\end{eqnarray}
There are certain key properties concerning the image of map $\Psi$ , i.e. the set of admissible local spectra, that are regarded as convexity theorems \cite{Brion,K84}. The convexity theorems are also central to the SLOCC classification \cite{sawicki}. The first convexity theorem is connected to the one-qubit marginal problem \cite{klyaczko} and ensures that $\Delta_\mathcal{H}=\Psi(\mathcal{H})$ is a convex polytope. This polytope is given by  the intersection of \cite{HSS03}
\begin{eqnarray}
\forall_{i}\ \frac{1}{2}-\lambda_{i}\leq\sum_{i\neq j}\left(\frac{1}{2}-\lambda_{i}\right),\,\label{nier1}
\end{eqnarray}
with the cube $\left[0,\frac{1}{2}\right]^{\times L}$ (for a different approach, see \cite{MT17}). Another important fact is that for the closure of any SLOCC class, i.e. for $\overline{\mathcal{C}}$ the corresponding (shifted) spectra of one-qubit RDMs forms a convex polytope $\Delta_{\mathcal{C}}=\Psi(\overline{\mathcal{C}})$ called the entanglement polytope (EP) \cite{Brion, walter, sawicki}. Obviously $\Delta_{\mathcal{C}}\subset \Delta_{\mathcal{H}}$. The method of entanglement polytopes is more robust then the standard approach of distinguishing all SLOCC classes. This is because for $L\geq 4$ the number of different SLOCC classes $\mathcal{C}$ is infinite, while the number of different polytopes $\Delta_{\mathcal{C}}$ is always finite \cite{ness, sawicki}. 

Finding entanglement polytopes is a difficult task as it requires knowing the generating set of covariants \cite{covariants,walter}. These are known only up to four qubits \cite{covariants}. Using this prior knowledge the entanglement polytopes for systems of up to four qubits have been found \cite{walter}. The important property of the entanglement polytopes is that they are typically not disjoint, $\Delta_\mathcal{C}\cap\Delta_{\mathcal{C}^\prime}\neq \emptyset$. For example, one easily shows that the entanglement polytope corresponding to the closure of the SLOCC class of the many-qubit $GHZ$ state is equal to $\Delta_{\mathcal{H}}$ and thus contains any polytope $\Delta_{\mathcal{C}}$. This motivates treating entanglement polytopes as entanglement witnesses, i.e for any $\phi\in \mathcal{H}$ we give a list of EPs that do not contain the RDM spectra of $\phi$. Obviously, the longer the list is, the better. Note that in experimental applications, only EPs of full dimension are useful. Therefore, we always restrict ourselves to considering only such polytopes. The decision-making  power of the set of EPs is thus determined by the volume of the region in $\Delta_{\mathcal{H}}$ where many fully-dimensional EPs overlap. States, whose one-qubit RDMs spectra belong to this region are poorly witnessed by entanglement polytopes. In this paper we study the witnessing power of EPs for the system of large number of qubits. As finding EPs, even for five qubits, is intractable, we use the connection between the EPs and the critical points of the linear entropy. Our main idea is to study the critical points via the random matrix theory for the Bernoulli and Wishart ensembles \cite{Gupta}. This way we show that the region in $\Delta_\mc{H}$ where EPs have a weak witnessing power always exists. We conjecture that in fact this region is large. The rigorous proof of this conjecture is beyond the methods used in this paper. We also point out that the procedure of entanglement distillation \cite{walter}, which in small systems is helpful in witnessing SLOCC classes whose polytopes intersect, is of limited usefulness when $L$ is large. We conclude that for large systems it is difficult to distinguish between generic entanglement classes using only single particle information. It is known that a generic quantum state is in a class that contains states, whose one-particle RDMs are maximally mixed. Equivalently, entanglement classes, whose entanglement polytopes do not contain $\overline 0$ form a set of measure zero in the space of pure quantum states. However, these classes have operational importance, as, for example, there is the class of $L$-qubit $W$-state among them. In this work we consider entanglement classes, whose EPs do not contain $\overline 0$. For entanglement classes of such a type, we show that the entanglement polytopes are concentrated in a distance $O(\frac{1}{\sqrt{L}})$ from zero. As we explain in the following section, this fact means that in large systems the locally maximally entangled states within a fixed entanglement class are usually close to states, whose one-qubit RDMs are maximally mixed. In this work, the locally maximally entangled states are the states that maximise the linear entropy (Eq.(\ref{lin-ent})) within a fixed entanglement class $\mc{C}$.


\section{Entanglement polytopes and the linear entropy}\label{sec:entropy}
The starting point for our considerations is establishing the connection between EPs and  critical points of the mean linear entropy, $E(\phi)$, of the one-qubit RDMs, 
\begin{eqnarray}\label{lin-ent}
E(\phi)=1-\frac{1}{L}\sum_{i=1}^L\tr\left(\rho_i(\phi)^2\right)=\frac{1}{2}-\frac{2}{L}\|\Psi(\phi)\|^2,
\end{eqnarray} 
where $\Psi(\phi)$ is determined by (\ref{Psi}) and $\|\Psi(\phi)\|^2=\sum_{i=1}^L \lambda_i^2(\phi)$. Up to three  qubits, the set of SLOCC entanglement classes is in bijection with the set of critical points of the linear entropy \cite{sawicki3}.  As a direct consequence of $(\ref{lin-ent})$ the value of $E(\phi)$ is determined by $\Psi(\phi)$, i.e. by spectra of one-qubit RDMs. Let $\mc{C}$ be an entanglement class and $\Delta_{\mathcal{C}}=\Psi(\overline{\mc{C}})$ its EP. EPs are convex and thus the restrictions of $E$ to $\Delta_\mc{C}$ has exactly two critical points, i.e. the minimum and the maximum. The minimum of $E$ is attained at $\overline{\lambda}_{Sep}=(\frac{1}{2},\ldots,\frac{1}{2})$. The states $\phi\in\overline{\mathcal{C}}$ that satisfy $\Psi(\phi)=\overline{\lambda}_{Sep}$ are all separable states. On the other hand, the maximum of $E$ is attained at the point that we denote by $\overline{\lambda}_{\mc{C}}\in\Delta_{\mc{C}}$ and call {\it the critical local spectra} of $\overline{\mc{C}}$. This point is the closest point of $\Delta_{\mc{C}}$ to the origin $\overline{0}$. Finally, we note that the maximal value of $E$ on $\overline{\mathcal{C}}$ which we denote by $E_{\mc{C}}$ is given by
\[
E_\mc{C}=\frac{1}{2}-\frac{2}{L}\|\overline{\lambda}_{\mc{C}}\|^2,
\]

Notably, the set of all possible critical local spectra, $\overline{\lambda}_{\mc{C}}$, is finite and can be found using a simple algorithm, where one considers convex combinations of points, that are the spectra of separable states that form a computational basis \cite{kirwan,MS15}. More precisely, let $\{\ket{i_1,\ldots,i_L}:i_j\in\{0,1\}\}$ be the computational basis  of $\mathcal{H}$. We have $\Psi(\ket{i_1,\dots,i_L})=\left((-1)^{i_1}\frac{1}{2},\dots,(-1)^{i_L}\frac{1}{2}\right)$. The algorithm is as follows. For every set of $L$ linearly independent vertices $\left(\overline v_1,\dots,\overline v_L\right)$ of the hypercube $\mathbb{H}_L$ centred at zero, whose vertices $\{\overline{v}_i\}$ have $\pm \frac{1}{2}$ coordinates, we:
\begin{enumerate}
\item Find $\overline\lambda_0$ - the closest to zero point of ${\rm Conv}\left(\overline v_1,\dots,\overline v_L\right)$.
\item If ${\lambda_0}_i\geq 0$ for all $i$ and $\overline\lambda_0$ does not belong to an edge of $\mathbb{H}_L$, then $\overline\lambda_0=\overline\lambda_\mc{C}$ for some SLOCC class $\mathcal{C}$.
\end{enumerate}
Adding $\overline {\lambda}_\mc{C}=\overline{0}$ and $\overline{\lambda}_{Sep}$ one obtains all possible $\overline{\lambda}_C$. The condition for the $L$-subset of vertices to be linearly independent assures that the corresponding polytope $\Delta_\mc{C}$ is of full dimension. The above procedure of finding the critical local spectra gives us a simple prescription for computing the critical values of $E_{\mc{C}}$, or equivalently of $\|\overline{\lambda}_\mc{C}\|^2$ -- the squared distances of the entanglement polytopes $\Delta_{\mc{C}}$ from zero. In the following we shall focus on the latter quantity. 

\section{Distribution of the distances of entanglement polytopes from zero}
\begin{figure}
\includegraphics{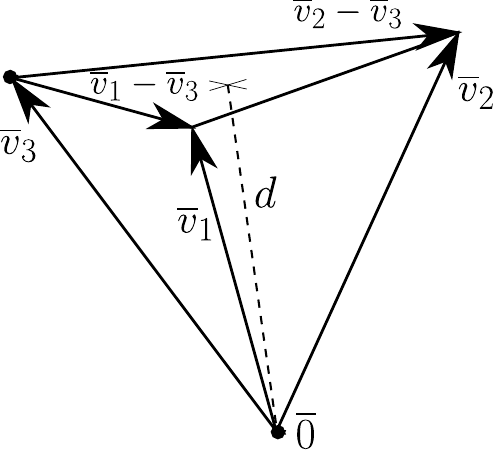}
\caption{An example of the simplex spanned at three vertices $\overline{v}_1$, $\overline{v}_2$ and $\overline{v}_3$ of $\mathbb{H}_3$ and the vertex $\overline{0}$.\label{fig:simplex}}
\end{figure}

\begin{figure*}
\includegraphics{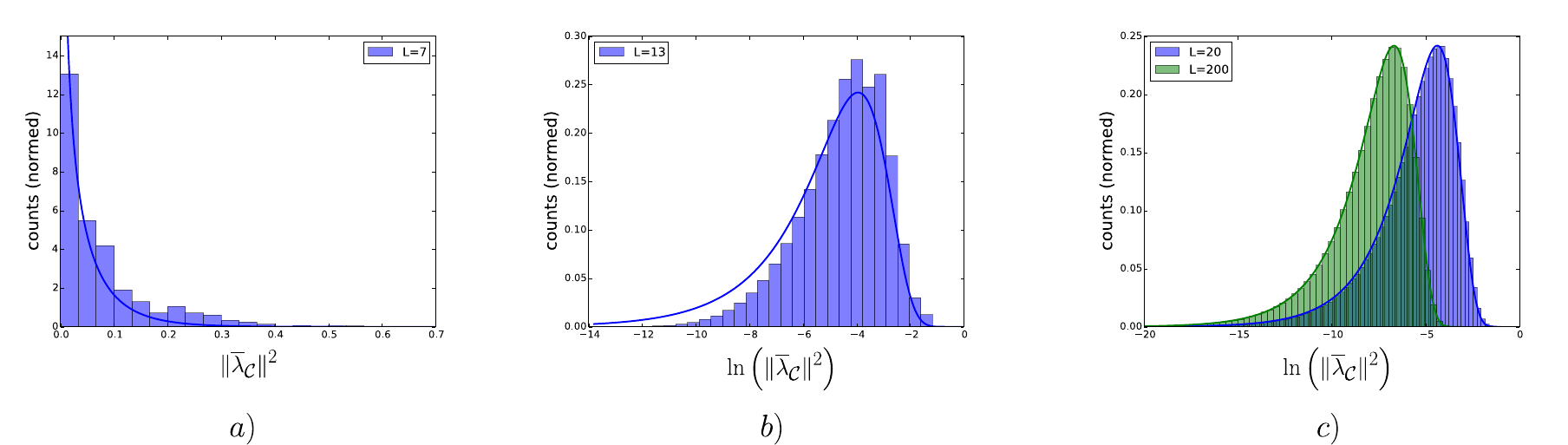}
\caption{(a) Distribution of $\|\overline{\lambda}_\mc{C}\|^2$ corresponding to fully-dimensional polytopes  for $7$ qubits. The plotted line is $Gamma\left(\frac{1}{2},2L\right)$. (b) and (c) Sample of $10^6$ points for the distrubition of $\ln \left (\|\overline{\lambda}_\mc{C}\|^2\right)$ for $13$, $20$ and $200$ qubits. The plotted lines are $LogGamma\left(\frac{1}{2},2L\right)$. For $L\geq20$ the distribution of $\|\overline{\lambda}_\mc{C}\|^2$ is well approximated by the $Gamma$ distribution.\label{fig:statistics}}. 
\end{figure*}

Firstly, we note that a subset of $L$ linearly independent vertices of $\mathbb{H}_L$ defines an $L$-simplex, which is spanned on those vertices and the vertex $\overline{0}$. The height of this simplex with respect to the vertex $\overline 0$ is equal to $d=\|\overline{\lambda}_{\mc C}\|$ for some $\mc{C}$ (see Figure \ref{fig:simplex}). In order to compute $d$, we will use the formula $V=\frac{1}{L}Ad$, where  $V$ is the volume of the simplex and $A$ is the area of the base. The volume is given by $V=\sqrt{\left|G(\overline v_1,\dots,\overline v_L)\right|}/L!$, where $G(v_1,\dots,v_L)$ is the Gramm matrix, whose entries read $G_{ij}=\overline v_i \cdot \overline v_j$ and $|G|$ denotes the determinant of $G$. The base of the simplex is spanned by vectors $\overline v_i-\overline v_L$, $i=1,2,\dots,L-1$, hence its area is given by $A=\sqrt{\left|G(\overline v_1-\overline v_L,\dots,\overline v_{L-1}- \overline v_L)\right|}/(L-1)!$. Thus 
\begin{equation}\label{height}
\|\overline{\lambda}_\mc{C}\|^2=\frac{\left|G(\overline v_1,\dots,\overline v_L)\right|}{\left|G(\overline v_1-\overline v_L,\dots,\overline v_{L-1}- \overline v_L)\right|}.
\end{equation}
For further calculation we rescale the distance and define
\begin{eqnarray}\label{dc}
d_\mc{C}^2:=4\|\overline{\lambda}_\mc{C}\|^2.
\end{eqnarray}
It is easy to see that $d^2_\mc{C}$ is given by (\ref{height}) with $\overline{v}_i$ having $\pm 1$ entries. The numerical computations that go through all $L$-subsets of vertices of $\mathbb{H}_L$ were carried out up to $L=7$. As one can see from Figure \ref{fig:statistics}a) values of $\|\overline{\lambda}_\mc{C}\|^2$ accumulate close to zero. In order to study this phenomenon in larger systems, we sampled $10^6$ $L$-subsets of vertices of $\mathbb{H}_L$ for $L$ up to $200$. The results for $\ln\left(\|\overline{\lambda}_\mc{C}\|^2\right )$ are plotted in Fig.~\ref{fig:statistics}b) and \ref{fig:statistics}c). We observe that the logarithmic distribution attains maximum at values less than zero and the maximum shifts in the negative direction when the number of qubits is increased. 
\section{Derivation of Gamma distribution}

In this section we describe the distribution of $\|\overline{\lambda}_\mc{C}\|^2$ using arguments from random matrix theory. Our main results is:

\paragraph*{\bf{Main result}}{\it The distribution of the squared distances of entanglement polytopes from the origin $\overline{0}$ for large $L$ tends to the gamma distribution
\begin{equation}\label{gamma_dist}
\|\overline{\lambda}_{C}\|^2\sim Gamma\left(\frac{1}{2},2L\right),
\end{equation}
where $Gamma(\alpha, \beta)$ is the probability distribution with the density:
\[
f(x)=\frac{\beta^\alpha}{\Gamma(\alpha)}x^{\alpha-1}e^{-\beta x}.
\]
}

Let us begin with the observation that vertices $\overline{v}_{i}$ (with $\pm 1$ coordinates) of the $L$-dimensional cube are uniformly distributed on $S^{L-1}$ with $r^2=L$. The distribution of $d^2_{\mc{C}}$ is obtained assuming every vertex $\overline{v}_i$ is given the same probability equal to $\frac{1}{2^L}$. Calculation of $d^2_\mc{C}$ for large $L$ is, however, difficult as the entries of $\overline{v}_i$ are discrete Bernoulli random variables. Our main idea is to do the calculations using continuous Gaussian random variables. As we will see later the distribution $d^2_\mc{C}$ calculated with Bernoulli random variables tends  for large $L$  to the distribution of  $d^2$, where $d^2$ is defined as in (\ref{dc}) and vectors  $\overline{v}_k=(v_1,\ldots,v_L)^t\in  \mathbb{R}^L$ are Gaussian vectors such that  $v_k\sim N(0,1)$ are independent Gaussian variables, i.e. $ \overline{v}\sim \frac{\exp\left(-\frac{1}{2}\|v\|^2\right)}{\sqrt{(2\pi)^{L}}}$. The distribution of $\overline {v}$ depends only on the length of $\overline v$ and thus is isotropic. Note that $\|\overline{v}\|^2$ is the sum of squares of $L$ independent Gaussian-distributed random variables, hence its distribution is the chi-squared distribution $\chi^2_L$ with the mean $L$ and the standard deviation $\sigma=\sqrt{2L}$ \cite{Gupta}. When $L\rightarrow \infty$ the ratio $\frac{\sqrt{2L}}{L}\rightarrow 0$, so vectors $\overline v_i$ are distributed on a thin shell. Thus, instead of calculating distribution of $d^2_\mc{C}$ using $\overline{v}_i$'s with $\pm 1$ entries, our aim is to calculate the analogous distribution for $\overline{v}_i\sim N(\overline{0},I)$. To this end, we start with a simpler calculation, namely the distribution of the ratio
\[
\frac{|G(\overline{v}_1,\ldots,\overline{v}_L)|}{|G(\overline{v}_1,\ldots,\overline{v}_{L-1})|},
\]
where $\overline{v}_i\sim N(\overline{0},I)$. Under our assumptions, matrix  $G:=G(\overline{v}_1,\ldots,\overline{v}_L)$ is a positive symmetric matrix distributed according to Wishart distribution $\mathcal{W}(L,I)$ \cite{Gupta}:
\[
\frac{1}{2^{L^2/2}\Gamma_L(\frac{L}{2})}|G|^{-\frac{1}{2}}e^{-(1/2)\mathrm{tr} G}.
\]
As $G$ is positive and symmetric we can decompose it using the Cholesky decomposition \cite{cholesky}, $G(\overline{v}_1,\ldots,\overline{v}_L)=TT^t$, where $T$ is a lower triangular matrix. The following theorem is a standard fact from the multivariate probability theory \cite{Gupta}:
\begin{fact}\label{thm1} (Bartlett) $T^2_{ii}$ are independent random variables  
\[
T_{ii}^2\sim Gamma\left(\frac{L-i+1}{2},\frac{1}{2}\right).
\]

\end{fact}
One can easily see that $T_{LL}^2$ is exactly the variable that we want to describe, i.e.
\[
T_{LL}^2=\frac{|G(\overline{v}_1,\ldots,\overline{v}_L)|}{|G(\overline{v}_1,\ldots,\overline{v}_{L-1})|}=\frac{|G|}{|G^{[L,L]}|}\sim Gamma\left(\frac{1}{2},\frac{1}{2}\right),
\]
where by $|G^{[L,L]}|$ we denote $(L,L)$-minor of the matrix $G$.

\subsection{Derivation of the $\|\overline{\lambda}_\mathcal{C}\|^2$-distribution using Gaussian vectors} 

In the following, we show that the distribution of $\frac{1}{4}d^2$ calculated using Gaussian vectors $\overline{v}_k\sim N(\overline{0},I)$ is exactly given by $Gamma\left(\frac{1}{2}, 2L\right)$. The mathematically rigorous proof showing that distribution $\frac{1}{4}d^2_\mc{C}$ calculated with Bernoulli vectors tends for large $L$ to $Gamma\left(\frac{1}{2}, 2L\right)$ will be published elsewhere. Here we only show numerically that $Gamma\left(\frac{1}{2}, 2L\right)$ matches histograms of $\frac{1}{4}d^2_\mc{C}$ almost perfectly already for $L\geq 20$ (Fig.~\ref{fig:statistics}). 

For calculation of the distribution of  $d^2$ we make the following observations. First, using antisymmetry of the determinant, we have that 
\[
|G(\overline{v}_1,\ldots,\overline{v}_{L-1},\overline{v}_L)|=|G(\overline{v}_1-\overline{v}_L,\ldots,\overline{v}_{L-1}-\overline{v}_L,\overline{v}_L)|
\]
Let us define 
\[
G^\prime:=G(\overline{v}_1-\overline{v}_L,\ldots,\overline{v}_{L-1}-\overline{v}_L,\overline{v}_L).
\]
This allows us to write 
\[
d^2=\frac{|G^\prime|}{|G^{\prime [L,L]}|}. 
\]
We note that $G^\prime=A^tGA$, where $A$ is the lower triangular matrix 
\[
A=\begin{pmatrix}
1 & 0 & \ldots & 0 &0 \\
0 & 1&\ldots &0&0  \\
\vdots & 0&\ddots&0 &0  \\
0 & 0&\ldots &1&0 \\
-1 &-1&\ldots &-1& -1
\end{pmatrix}
\]
and $G=G(\overline{v}_1,\ldots, \overline{v}_L)$ with $\overline{v}_k \sim N(\overline{0},I)$. The following fact is crucial for our purposes \cite{Gupta}.

\begin{fact}\label{thm2}
 Assume $G\sim \mathcal{W}(L,I)$ and $A$ is an invertible matrix. Then $G^\prime = A^tGA\sim \mathcal{W}(L,\Sigma)$, where $\Sigma:=A^tA$. In other words $G^\prime$ is distributed according to the Wishart distribution $\mathcal{W}(L,\Sigma )$, that is, $G^\prime$ is a Gramm matrix of $L$ sample vectors $\overline{w}_i\sim N(\overline{0},\Sigma)$
 \[
\overline{w}\sim \frac{\exp\left(-\frac{1}{2}\overline{w}^t\Sigma^{-1}\overline{w}\right)}{\sqrt{(2\pi)^{L}\det\Sigma}}
\]
\end{fact}

As a direct consequence, we get that for vectors $\overline{v}_i\sim N(\overline{0},I)$ and vectors $\overline{w}_i\sim N(\overline{0},\Sigma)$  the distributions of 
\[
\frac{|G(\overline{v}_1,\ldots,\overline{v}_L)|}{|G(\overline{v}_1-\overline{v}_L,\ldots,\overline{v}_{L-1}-\overline{v}_L)}\,\,\mathrm{and}\,\,\frac{|G(\overline{w}_1,\ldots,\overline{w}_L)|}{|G(\overline{w}_1,\ldots,\overline{w}_{L-1})|} 
\]
are the same. Therefore, we need to calculate the distribution of $\frac{|G(\overline{w}_1,\ldots,\overline{w}_L)|}{|G(\overline{w}_1,\ldots,\overline{w}_{L-1})|} $ for $\overline{w}_i\sim N(0,\Sigma)$. To this end, let $RR^t$ be the Cholesky decomposition of $\Sigma=A^tA$ and let $TT^t$ be the Cholesky decomposition of $G\sim \mathcal{W}(L,I)$. Using Fact \ref{thm2} one easily see that $RTT^tR^t\sim\mathcal{W}(L,\Sigma)$.  Moreover matrices $R$ and $T$ are lower triangular and their product is also a lower triangular matrix. Therefore the ratio $\frac{|G(\overline{w}_1,\ldots,\overline{w}_L)|}{|G(\overline{w}_1,\ldots,\overline{w}_{L-1})|}$ has the same distribution as
\[
(RT)^2_{LL}=R^2_{LL}T^2_{LL}=\frac{1}{L}T_{LL}^2.
\]
By Fact \ref{thm1} we know that $T_{LL}^2\sim Gamma\left(\frac{1}{2},\frac{1}{2}\right)$. The main result is obtained using a property of gamma distributions, which states that if a random variable $X$ is distributed according to $Gamma(\alpha,\beta)$, then $\lambda X$ has distribution $Gamma(\alpha,\frac{\beta}{\lambda})$. In our case, $\alpha=\beta=\frac{1}{2}$ and $\lambda=\frac{1}{4L}$. Fig.~\ref{fig:statistics} shows that histograms of $\|\lambda_\mc{C}\|^2=\frac{1}{4}d^2_\mc{C}$ fit very well $Gamma\left(\frac{1}{2},2L\right)$.

\section{Conclusions}  
\begin{figure}[h]
\includegraphics{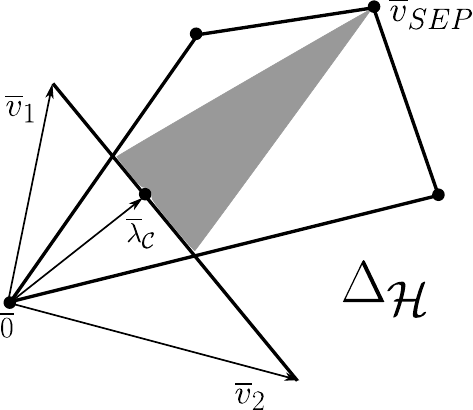}
\caption{For each $\overline\lambda_\mc{C}$, one can construct a thin polytope contained in $\Delta_\mc{C}$, whose basis contains $\overline\lambda_\mc{C}$ and whose one vertex is $\overline v_{SEP}$ (grey area). The intersection of such polytopes over all classes $\mc{C}$ gives the area of weak entanglement witnessing power.}
\label{thin}
\end{figure}

One of the corollaries from the main result is existence of a region in $\Delta_\mc{H}$, where the witnessing power of entanglement polytopes is weak. Namely, consider polytopes, that do not contain $\overline 0$ and are of full dimension. The lower facet of such a polytope, $\Delta_\mc{C}$, is given by the intersection of the hyperplane perpendicular to $\overline\lambda_\mc{C}$ and containing $\overline\lambda_\mc{C}$ with $\Delta_\mc{H}$ \cite{ness}. Moreover, because the set of separable states is in the closure of every entanglement class, we have $\left(\frac{1}{2},\dots,\frac{1}{2}\right)\in\Delta_\mc{C}$. By the convexity of entanglement polytopes, the convex hull of the lower facet and the separable vertex is contained in $\Delta_\mc{C}$ (see Fig.\ref{thin}).
Constructing such a subpolytope for each entanglement class, we obtain the set of cones, whose lower facets are accumulated in the distance of $\frac{1}{2\sqrt{L}}$ from zero and that share the separable vertex. Therefore, they intersect in a region concentrated along the line from zero to the separable vertex. The numerical data that we obtained for $7$ qubits suggests that the entanglement polytopes are in fact larger than the subpolytopes described above. Namely, by constructing states that have the critical spectra $\overline\lambda_\mc{C}$ and acting on such states only with diagonal SLOCC operations, one can reach more vertices of $\Delta_\mc{H}$ than just the separable vertex.

Another conclusion from the main result of this work is the fact that the optimal entanglement distillation protocol, which has been proposed in \cite{walter} as a way of improving the witnessing power of entanglement polytopes, does not work for large number of qubits. Such a protocol transforms a given state $\phi\in\mc{C}$ using SLOCC operations to a state with critical local spectra $\overline{\lambda}_\mc{C}$. This is done by following the gradient flow of $E$. The gradient $\mathrm{grad}E(\psi)$ is always tangent to $\mc{C}_\phi$ at point $\phi$ \cite{ness}, hence the lines of gradient flow can be realised by SLOCC operations. A similar protocol using random SLOCC operations has been realised experimentally in \cite{exp2}. The main idea behind using the entanglement distillation protocols is the fact that for small systems the critical spectra $\overline{\lambda}_\mc{C}$ usually belong to the intersection of a small number of polytopes. For a system of $3$ qubits, for example, such a procedure allows one to perfectly decide whether a given state has the type of entanglement of the GHZ- or of the W-state. However, when $L$ is large, the critical spectra are very densely packed in a region close to $\overline{0}$, so entanglement distillation is usually not helpful.

\begin{figure}[h]
\includegraphics{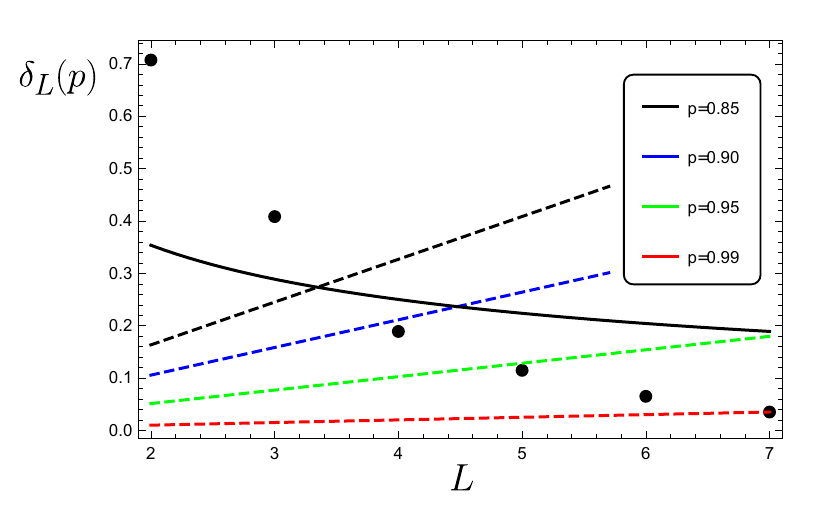}
\caption{Precision $\delta_L(p)$ required for distinguishing between generic polytopes in large $L$ limit (black line) vs. precision required for distinguishing between all polytopes (black dots). In order to distinguish between all polytopes for $L=7$, the required purity is around $0.99$. However, a generic polytope requires purity around $0.95$.}
\label{purity}
\end{figure}

Our results also give a quantitative estimation of the required purity of states in experimental applications of the entanglement polytopes' method. A calculation based on perturbation theory shows that for a mixed state $\rho$ with $\tr\rho^2=p$ there exists a pure state $\psi$ such that $\bra{\psi}\rho\ket{\psi}\geq p$ and whose vector of local spectra differs from the local spectra of $\rho$ by at most $\delta_L(p)=\frac{L}{2}(1-\sqrt{2p-1})$ \cite{walter}. This means that the entanglement polytopes can witness entanglement successfully with precision $\delta_L(p)$ \cite{walter,exp1}. Hence, for a generic entanglement polytope the required precision is $\delta_L(p)\leq\frac{1}{2\sqrt{L}}$. However, in order to be able to distinguish the polytopes perfectly, we need the precision which is less than the smallest value of $\|\overline\lambda_\mc{C}\|$ of all entanglement classes. These quantities are compared to $\delta_L(p)$ on Fig.\ref{purity}.

The techniques introduced in this paper can be generalised and used to formulate a general framework for studying multipartite entanglement in systems of many qudits, bosons or fermions in the limit of many particles. Namely, one can construct a similar algorithm for finding the critical values of the linear entropy by looking at the local spectra of states that form a computational basis of the proper Hilbert space. By sampling the sets of local spectra of basis states and computing volumes of the corresponding simplices, one constructs different random matrix models, where the limit or large matrix size corresponds to the limit of a large number of particles.

\begin{acknowledgments}
This work was supported by National Science Centre, Poland under the grant SONATA BIS: 2015/18/E/ST1/00200 and Polish Ministry of Science and Higher Education ``Diamentowy Grant'' no. DI2013 016543.
\end{acknowledgments}

\bibliographystyle{apsrev4-1}

\begin{thebibliography}{9}

\bibitem{Horodecki} Horodecki, R., Horodecki, P., Horodecki, M., Horodecki, K., Quantum entanglement Rev. Mod. Phys. Vol. 81, No. 2, pp. 865-942, 2009

\bibitem{Nilsen} M. A. Nielsen, Phys. Rev. Lett. 83, 436 - 439, (1999).

\bibitem{KrausLOCC} J. I. de Vicente, C. Spee, and B. Kraus Phys. Rev. Lett. 111, 110502 (2013)

\bibitem{Dur}W. D\"ur, G. Vidal, and J. I. Cirac, Phys. Rev. A 62, 062314 (2000).

\bibitem{Verstraete} F. Verstraete, J. Dehaene, B. De Moor, and H. Verschelde, Phys. Rev. A 65,
052112 (2002).

\bibitem{Levay}P. L\'evay, and P. Vrana, Phys. Rev. A 78, 022329, (2008).

\bibitem{kus} J. Schliemann, J. I. Cirac, M. Ku\'s, M. Lewenstein,  D. Loss, Phys. Rev. A 64, 022303, (2001).

\bibitem{holweck} F. Holweck, P. L\'evay, J. Phys A: Math and Theo 49 (8), 085201, (2016).

\bibitem{wallach1} G. Gour, N. R. Wallach, J. Math. Phys. 51, 112201, (2010).

\bibitem{Dokovic} O. Chterental and D. \u{Z}. Dokovi\'c, Normal forms and tensor
ranks of pure states of four qubits, in Linear Algebra Research
Advances (Nova Science Publishers, Inc.) pp. 133-167, (2007)

\bibitem{wallach2}N. R. Wallach, Quantum computing and entanglement for mathematicians. Representation theory and complex analysis, 345-376, Lecture Notes in Math., 1931, Springer, Berlin, (2008)

\bibitem{turner} J. Turner and J. Marton, SIGMA 13, 028, (2017)

\bibitem{kirwan} F. C. Kirwan, {\it Cohomology of Quotients in Symplectic and Algebraic Geometry}, Mathematical Notes, Vol. 31, Princeton Univ. Press, Princeton, 1984

\bibitem{ness} L. Ness, Amer. J. Math. 106(6), 1281-1329, 1984

\bibitem{MS15} T. Maci\k{a}\.{z}ek, A. Sawicki,  J. Phys. A: Math. Theor. 48 045305, DOI: 10.1088/1751-8113/48/4/045305, (2015)

\bibitem{Brion} M. Brion, in S\'eminaire d'Alg\'ebre P. Dubreil et M.-P. Malliavin, Springer, p. 177,1987

\bibitem{klyaczko}A. Klyachko, J. Phys.: Conf. Ser. 36, 72 (2006).

\bibitem{klyaczko2} A. Klyachko, arXiv:quant-ph/0206012 (2002).

\bibitem{Gupta} A. K. Gupta, D. K. Nagar, Matrix Variate Distributions, Champman and Hall/CRC, Boca Raton, 2000


\bibitem{walter}M. Walter et. al., Science 340 (6137), 1205-1208, 2013

\bibitem{sawicki} A. Sawicki, M. Oszmaniec, M. Ku\'s,  Rev. Math. Phys. 26, 1450004, 2014

\bibitem{sawicki2}A. Sawicki, A, Huckleberry, M Ku\'s, Commun. Math. Phys. 305 (2), 441-468, 2011

\bibitem{symplecticreview}A. Sawicki et. al., arXiv:1701.03536, 2017

\bibitem{SK11} A. Sawicki, M. Ku\'s, J. Phys. A: Math. Theor. 44 495301, 2011
\bibitem{SWK13} A. Sawicki, M. Walter, M. Ku\'s, J. Phys. A 46, 055304, 2013
\bibitem{CCCQ} M. Oszmaniec, P. Suwara, A. Sawicki, J. Math. Phys. 55, 062204, 2014

\bibitem{ST13} A. Sawicki, V. V. Tsanov, J. Phys. A: Math. Theor. 46 265301, 2013

\bibitem{HKS13} A. Huckleberry, M. Ku\'s, A. Sawicki, J. Math. Phys. 54, 022202, 2013

\bibitem{walter2}M. Christandl, B. Doran, S. Kousidis, M. Walter, Commun. Math. Phys. 332, 1-52 (2014)

\bibitem{maciazek1}T. Maci\k{a}\.zek, M. Oszmaniec, A. Sawicki, J. Math. Phys. 54, 092201 (2013)

\bibitem{HSS03}A. Higuchi, A. Sudbery, J. Szulc, Phys. Rev. Lett. 90, 107902, (2003)

\bibitem{covariants}E. Briand, J.-G. Luque, and J.-Y. Thibon,  J. Phys. A 38, 9915 (2003).

\bibitem{K84} F. C. Kirwan, Invent. Math. 77, 547552. (1984)

\bibitem{sawicki3} A. Sawicki, M. Oszmaniec, M Ku\'s, Physical Review A 86 (4), 040304, (2012)

\bibitem{cholesky} G. H. Golub, C. F. Van Loan, {\it Matrix Computations (3rd ed.)}, Baltimore: Johns Hopkins, ISBN 978-0-8018-5414-9. (1996)

\bibitem{exp1} G. H. Aguilar, S. P. Walborn, P. H. Ribeiro, L. C. Céleri, Phys. Rev. X 5, 031042 (2015)

\bibitem{exp2} Yuan-Yuan Z., M. Grassl, Bei Z., Guo-Yong X., Chao Z., Chuan-Feng Li, Guang-Can G., npj Quantum Information 3, Article number: 11, doi:10.1038/s41534-017-0007-5. (2017) 

\bibitem{MT17} T. Maci\k{a}\.{z}ek, V. Tsanov, {\it Quantum marginals from pure doubly excited states}, J. Phys. A: Math. Theor. 50 465304 (2017)

\bibitem{tennie2016pinning1} Tennie, F. and Ebler, D. and Vedral, V. and Schilling, C., {\it Pinning of fermionic occupation numbers: General concepts and one spatial dimension}, Phys. Rev. A, 93(4), 042126, 2016

 \bibitem{tennie2016pinning2} Tennie, F. and Vedral, V. and Schilling, C., {\it Pinning of fermionic occupation numbers: Higher spatial dimensions and spin}, Phys. Rev. A, 94(1), 012120, 2016
  
\bibitem{tennie2017influence} Tennie, F. and Vedral, V. and Schilling, C., {\it Influence of the fermionic exchange symmetry beyond {P}auli's exclusion principle}, Phys. Rev. A, 95(2), 022336, 2017

\bibitem{schilling2013natural} Schilling, C., {\it Natural orbitals and occupation numbers for harmonium: Fermions versus bosons}, Phys. Rev. A, vol. 88, no 4, pp 042105, 2013

\bibitem{schilling2015hubbard} Schilling, C., {\it Hubbard model: Pinning of occupation numbers and role of symmetries}, Phys. Rev. B 92, 155149, 2015

\bibitem{benavidespinning2} Benavides-Riveros, Carlos L. and Springborg, Michael, {\it Quasipinning and selection rules for excitations in atoms and molecules}, Phys. Rev. A 92, 012512, 2015

\bibitem{S15} Schilling, C., {\it Quasipinning and its relevance for N-fermion quantum states}, Phys. Rev. A 91, 022105 (2015)

\bibitem{OS17} Legeza, \''{O}., Schilling, C., {\it Role of the pair potential for the saturation of generalized Pauli constraints}, arXiv:1711.09099 (2017)
\end{thebibliography}

\end{document}